\documentclass[aps,superscriptaddress,floatfix,nofootinbib,preprintnumbers,eqsecnum,twocolumn,dvipsnames]{revtex4-2}
\counterwithout{equation}{section}
\usepackage{amsmath,float,graphicx,placeins,amssymb,multirow,pgfplots,pgfplotstable}
\usepackage{empheq}
\usepackage{tikz}
\usepackage{comment}
\usepackage{enumerate,slashed,cancel,comment,color,bbm,graphicx,rotating,placeins,mathtools,multirow,hhline}
\usepackage[normalem]{ulem}
\usepackage{array}
\usepackage{hyperref}
\usepackage{color}
 \hypersetup
 {
   colorlinks,
   linkcolor={blue!80!black},
   citecolor={blue!70!black},
   urlcolor={blue!70!black}
 }

\usepackage{hyperref}

\newcommand{\be}{\begin{equation}}
\newcommand{\ee}{\end{equation}}
\newcommand{\bea}{\begin{eqnarray}}
\newcommand{\eea}{\end{eqnarray}}
\newcommand{\ba}{\begin{aligned}}
\newcommand{\ea}{\end{aligned}}

\begin{document}
\sloppy

\title{Primordial Black Hole Hot Spots and Nucleosynthesis}

\begin{abstract} 
Upon their evaporation via Hawking radiation, primordial black holes (PBHs) deposit energy in the ambient plasma on scales that can be smaller than the typical distance between two black holes, leading to the formation of hot spots around them. We investigate how such hot spots may act as a shield against the release of low-energy photons, affecting PBH's capacity to dissociate light nuclei after Big-Bang Nucleosynthesis through photo-dissociation. We find that this shielding effect is particularly relevant for PBHs with masses between $10^{11}$g and $4\times 10^{12}$g. We emphasize that the magnitude of this effect is highly dependent on the specific shape of the temperature profile around PBHs and its time evolution, stressing the necessity for a comprehensive study of PBH hot spots and their dynamics in the future.
\end{abstract}

\author{Clelia Altomonte}
\email[Email address: ]{clelia.altomonte@kcl.ac.uk}

\author{Malcolm Fairbairn}
\email[Email address: ]{malcolm.fairbairn@kcl.ac.uk}

\author{Lucien Heurtier}
\email[Email address: ]{lucien.heurtier@kcl.ac.uk}
\affiliation{Theoretical Particle Physics and Cosmology, King’s College London,\\ Strand, London WC2R 2LS, United Kingdom}

\maketitle

\section{Introduction}

Primordial Black Holes (PBHs) are fascinating objects that may be candidates for the missing dark matter in the Universe~\cite{Carr:2020gox, Green:2020jor} and originate from the collapse of primordial curvature over-densities \cite{Carr:1975qj}, that can originate from various physical mecanisms~\cite{Carr:1993aq,Ivanov:1994pa,Garriga:2015fdk,Deng:2016vzb,Deng:2017uwc,Kusenko:2020pcg,Heurtier:2022rhf, Ai:2024cka, Maeda:1981gw,Sato:1981gv,Kodama:1981gu,Kodama:1982sf,Hsu:1990fg,Liu:2021svg,Kawana:2022olo,Gouttenoire:2023naa,Hawking:1982ga,Moss:1994iq,Moss:1994pi, Khlopov:1998nm, Crawford:1982yz,Gross:2021qgx,Baker:2021sno,Kawana:2021tde, Dolgov:1992pu,Dolgov:2008wu,Kitajima:2020kig,Kasai:2022vhq,Martin:2019nuw,Martin:2020fgl,Rubin:2000dq, Vachaspati:2017hjw,Ferrer:2018uiu,Gelmini:2022nim,Gelmini:2023ngs,Hawking:1987bn,Polnarev:1988dh,Fort:1993zb,
Garriga:1993gj,Caldwell:1995fu,MacGibbon:1997pu,Jenkins:2020ctp,Blanco-Pillado:2021klh}.  Nevertheless, PBHs with masses less than $10^{15}$ g evaporate on time scales smaller than the age of the Universe, due to Hawking radiation, so they are not expected to be around today. 

One of the main observational constraints on evaporating PBHs comes from the effect of their Hawking evaporation after the onset of Big Bang Nucleosynthesis (BBN)~\cite{Sarkar_1996, Jedamzik_2009,Izotov_2014,Carr_2010, Carr:2020gox, Forestell_2019}. 
To derive such constraints, previous works generally assumed that the flux of particles emitted by evaporating PBHs is homogeneous in space. This assumption relies on the observation that the average distance between PBHs at the time of their evaporation is: $(i)$ much smaller than the Hubble radius--meaning that there are many PBHs per Hubble patch when they evaporate--and $(ii)$ much smaller than the typical mean free path of the low-energy particles in play in the dissociation of light nuclei formed during BBN. Within this framework, PBHs are assumed to radiate particles locally, but these particles quickly lose energy after scattering on the homogeneous plasma. It is only after this {\em reprocessing} of the Hawking radiation to lower energy that particles emitted by many PBHs propagate freely until they dissociate light nuclei in a homogeneous manner. {However, it is during this reprocessing that local variations of the plasma temperature may play a significant role.} 

In this paper, we challenge the assumption of homogeneity by accounting for the fact that the local plasma around PBHs may take the form of a hot spot, as suggested in Refs.~\cite{He:2022wwy,He:2024wvt,Das:2021wei}. As we will show, a local increase in the temperature around PBHs can lead to temporary trapping of the particles radiated, effectively reducing the capacity of PBHs to disturb BBN. For simplicity, we will consider in this paper only the case of PBHs in the mass range $10^{11}\mathrm{g}\lesssim M\lesssim 10^{13}\mathrm{g}$ for which the most relevant constraint comes from photodissociation\footnote{Localisation effects were mentioned in studies like~\cite{Acharya_2022}, and shown to be important to the phenomenology of evaporating PBHs in the context of black hole superradiance~\cite{Pani_2013,Chluba_2015} or PBH accretion~\cite{Ali_Ha_moud_2017} for heavier PBHs.}.

Photodissociation is the nuclear reaction by which a nucleus decays  after absorbing a photon of energy larger than its nuclear binding energy~\cite{Luo_2021, Kawasaki:1994sc}. 
This process is relevant for BBN constraints since it can change the relative abundances of primordial light elements. For instance, the breaking up of helium nuclei can reduce the helium abundance while increasing the primordial deuterium abundance\cite{Keith_2020}. For a summary of the primary photodissociation reactions, see table 6 from Ref. \cite{Luo_2021}. 
Here we report, for illustrative purposes, one of the possible photodissociation processes for helium, ${ }^3 \mathrm{He}(\gamma, p)^2 \mathrm{H}$, whose  reaction is 
\begin{equation}
{ }^3 \mathrm{He}+\gamma \longrightarrow {}^2\mathrm{H} + p
\end{equation}
and threshold energy of about 5.5 Mev.

While the threshold energy for photodissociating a given element species changes depending on the atomic element, the probability of photodissociation also depends on the temperature of the plasma. Indeed, when considering the photodissociation of helium nuclei, it is assumed that the temperature of the background radiation needs to be lower than that of $e^{+} e^{-}$ pair production ($T \sim 0.4 \mathrm{keV}$) \cite{Keith_2020, Kawasaki_1995}, otherwise the photodissociating photons would be absorbed by the plasma before they can break up helium nuclei\cite{Keith_2020}. More generally, the mean-free-path of photons before the plasma absorbs them through electron-positron annihilation needs to be larger than the mean distance between the atoms to be photodissociated, which in turn is related to the element species density. For this reason, when deriving BBN constraints on PBHs based on photodissociation, it is necessary to evolve the Hawking radiated photon spectrum \cite{Acharya_2020, Lucca_2020,Poulin_2015}, taking into account both its interaction with the surrounding plasma, the local plasma temperature and the universe expansion rate \cite{Kuznetsova_2010, carmona2022modificationmeanfreepath, Kawasaki_1995}.

An additional necessary ingredient for this calculation is the modelling of the PBHs energy and particle injection into the plasma. The two main approaches used are the "on-the-spot" (i.e. instantaneous) approximation of the energy injection due to PBHs \cite{St_cker_2018}, and methods where energy injection and deposition are considered over various deposition channels and at different times (cf. e.g., refs. \cite{poulter2019cmbconstraintsultralightprimordial}\cite{Lucca_2020}). 
As noted in the introduction, all of the mentioned processes and the plasma are generally assumed to be isotropic and homogeneous. 

\vspace{5pt}
\section{Primordial Black Holes}
In this paper, We consider the case of a monochromatic mass distribution\footnote{Note that in general, PBHs can have extended mass distributions, see e.g.~\cite{Sasaki:2018dmp, Carr:2020xqk, Cheek:2022mmy}.} forming from the collapse of super-horizon perturbations re-entering the horizon during the radiation-dominated era. We denote the temperature of the Universe at formation as $T_i$. Their mass is related to the Hubble horizon at formation $H_i$ via
\begin{align}\label{eq:initial_mass}
 M_i &= \frac{4\pi}{3}\, \gamma\, \frac{\rho_{\rm rad}(T_i)}{H_i^3}\sim 2\cdot10^5{\rm g} \left(\frac{10^{13}~{\rm GeV}}{T_i}\right)^2\!\!\!\!\!\,,
\end{align}
where $\gamma \sim 0.2$ is the gravitational collapse factor for a radiation-dominated Universe and $\rho_{\rm rad}$ the radiation energy density. The relative abundance of PBHs at formation, related to the energy density of PBHs, $\rho_{\rm PBH}^i$, is usually quantified~\cite{Carr:2020gox} using the parameter
\begin{align}\label{eq:beta}
 \beta^\prime \equiv \gamma^{1/2}\left(\frac{g_{\star}(T_i))}{106.75}\right)^{1/4}\frac{\rho_{\rm BH}^i}{\rho_{\rm rad}(T_i)}\,,
\end{align}
where $g_{\star}(T_i)$ denotes the number of relativistic degrees of freedom at formation. Considering a semi-classical approach, Hawking~\cite{Hawking:1975vcx} showed that a PBH of mass $M$ emits all existing particles like a blackbody with temperature
$T_H = (8\pi\, G M)^{-1}\,\sim 10^4~{\rm GeV} (10^9 g /M)$.
After the Universe becomes cooler than the PBHs' initial temperature, Hawking radiation becomes efficient, and radiates each particle species $\ell$ with an emission rate per energy and time , $d^{2}N_\ell/(dtdE)$. These rates contribute collectively 
to the PBH mass loss rate, given by~\cite{Hawking:1975vcx, MacGibbon:1990zk, MacGibbon:1991tj, Cheek:2021odj}
\begin{align} \label{eq:MEq}
 \frac{dM}{dt} = -\sum_\ell \int \frac{d^{2}N_\ell}{dt\, dE}\, E\, dE \equiv - \varepsilon(M)\, \frac{M_p^4}{M^2}\,,
\end{align}
where $\varepsilon(M)$ denotes the total effective evaporation function. 
Because the overall scaling of a PBH evaporation rate is $\dot M/M\sim M^{-3}$, it is remarkable that the evaporation of PBHs in cosmology is an accelerating process. Indeed, the smaller the PBH, the hotter its Hawking temperature and the larger its emission rate. The initial evaporation rate at the PBH formation (when its mass $M_i$ is the largest), $\Gamma_{\rm ev} = \varepsilon(M_i)M_p^4/M_i^3$ thus sets the overall PBH evaporation time $t_{\rm ev}\sim \Gamma_{\rm ev}^{-1}$ to a good accuracy. This happens when the Universe's temperature equals $T_{\rm ev}\equiv (90/8\pi^3 g_\star(T_ev))^{1/4}\sqrt{\Gamma_{\rm ev}M_p}$.
This temperature can be assumed to remain constant until evaporation is over unless PBHs are sufficiently abundant to reheat the Universe when they evaporate. In what follows, we will consider PBH abundances that are small enough to have a negligible impact on the Universe's temperature far away from the PBH horizon. However, as we will see in the next section, this conclusion does not hold for the radiation plasma directly surrounding PBHs during their evaporation, as the acceleration of the evaporation and the quick increase of the PBHs' Hawking temperature lead to a local temperature gradient around PBHs.

\begin{figure*}
    \centering
    \includegraphics[width=0.33\linewidth]{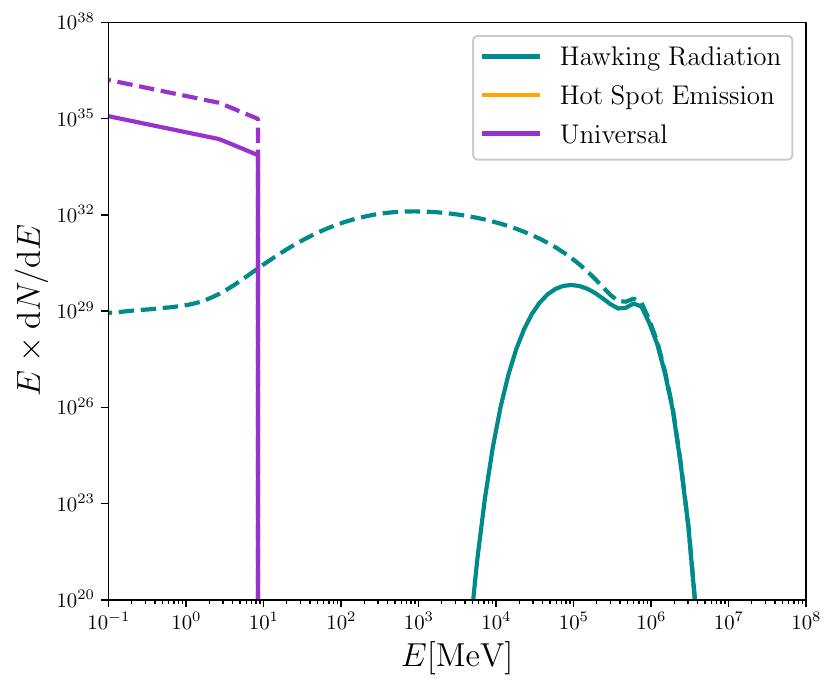}\includegraphics[width=0.33\linewidth]{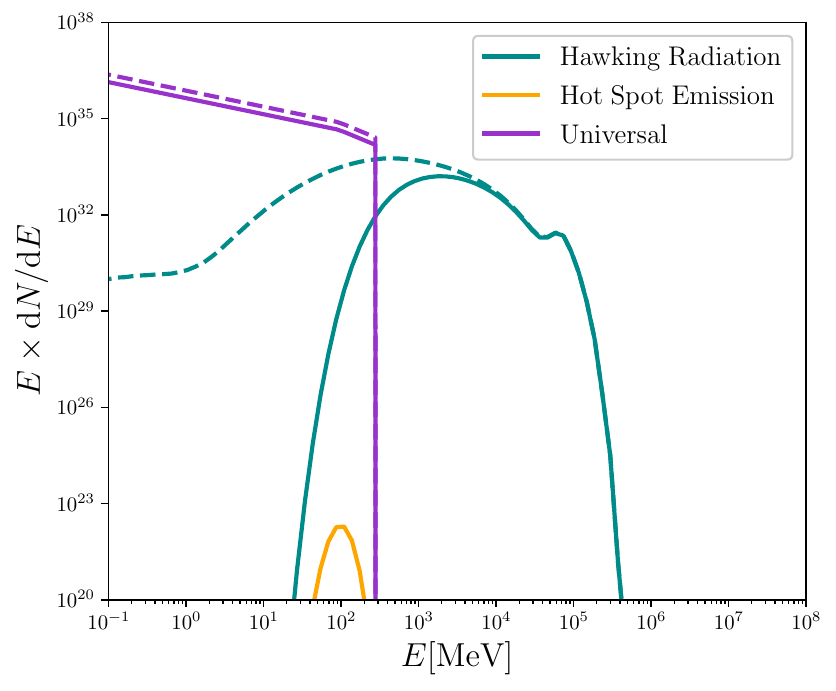}\includegraphics[width=0.33\linewidth]{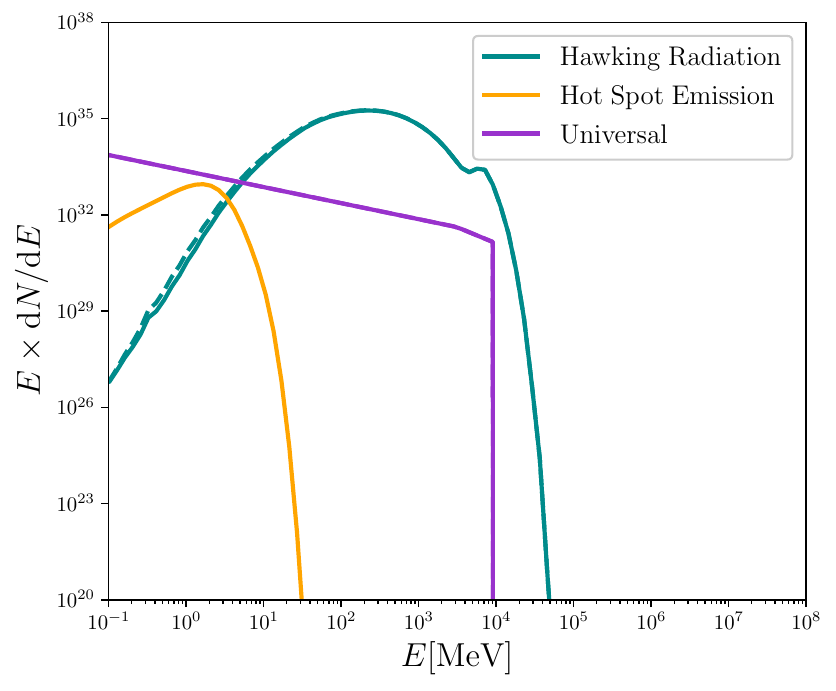}\caption{\label{fig:fluxBlackHawk}\footnotesize Different contributions to the photon energy distributions, for PBHs with  masses $10^{11}\mathrm{g}$ (left), $10^{12}\mathrm{g}$ (center), and $10^{13}\mathrm{g}$ (right), when considering (ignoring) the presence of a hot spot are depicted using plain (dashed) lines.}    
\end{figure*}
\vspace{5pt}
\section{Hot spot dynamics\label{sec:hotspot}}
When they evaporate, PBHs do not homogeneously reheat the Universe. Instead, they deposit their energy locally, forming hot spots around them with a characteristic radial temperature profile. The qualitative shape and time evolution of such a profile was first considered in Ref.~\cite{Das:2021wei} and refined--accounting for the Landau-Pomeranchuk-Migdal (LPM) effect in the calculation of the Hawking radiation energy deposition rate in Ref.~\cite{He:2022wwy}. These results were later supported by a Boltzmann-improved numerical treatment~\cite{He:2024wvt} that confirmed the validity of the qualitative shape of the radial temperature profile derived in Ref.~\cite{He:2022wwy} in the region where the assumption of {\em local thermal equilibrium} can be trusted. In these references, derivations relied on a few important assumptions: 
\begin{itemize}
    \item The plasma is assumed to be in local thermal equilibrium, such that it can be uniquely described at every radius $r$ by a scalar function $T(r)$;
    \item The energy deposition rate of high-energy Hawking radiation particles with momentum $k$ onto a significantly lower-energy plasma of temperature $T\ll k$ is estimated, accounting for the Landau–Pomeranchuk–Migdal (LPM), using the ansatz~\cite{He:2022wwy}
    $\Gamma_{\rm dep} \sim \alpha^2T\sqrt{T/k}$;
    \item The diffusion rate that encodes the radial cooling of the profile is of the Bethe-Heitler form $
    \Gamma_{\rm diff}\sim \alpha^2 T$,
\end{itemize}
where in these two expressions, the coupling constant $\alpha \sim 0.1$ is assumed to be universal to all particle interactions for simplicity. In App. A, we report the specific temperature profile $T(r)$ obtained in these studies and that we will use throughout this work.
Note that once the temperature in the profile drops below the electron mass $m_e$, the plasma is made mainly of inert photons that cannot sustain local thermal equilibrium. Elucidating how the photon plasma evolves under the effect of Hawking evaporation for temperatures below this threshold requires an increased level of sophistication compared to Refs.~\cite{He:2022wwy, He:2024wvt} which we let for future work. Instead, we will consider in what follows that we can trust the qualitative shape of the profile as described in Eq.~\eqref{eq:profile}, as long as $T(r)>\max(T_{\rm ev},m_e)$, and consider that $T(r)=T_{\rm ev}$ otherwise. 

The presence of a hot spot can affect the PBH capacity for photo-dissociating light nuclei formed during BBN in two ways: First, the large temperature in the vicinity of the BH compared to the background temperature of the Universe can shield part of the Hawking emission by decreasing locally the mean free path of particles radiated by the BH. Second, the hot spot itself can reprocess the energy of Hawking-radiated particles and radiate low-energy photons at a different rate than what is expected from the direct Hawking emission. We now estimate these two effects.
\vspace{5pt}
\section{Hot Spot Shielding}
After elementary particles are radiated with an energy comparable to the PBH Hawking temperature, they quickly shower to produce large amounts of secondary photons at much lower energy. In the absence of a hot spot, such photons scatter throughout the Universe's plasma, slowly losing energy, and typically propagate over distances much larger than the separation between two PBHs. In the presence of a hot spot, local temperature variations can significantly alter the photon mean free path, and not all photons produced by Hawking radiation can escape. 

Following Ref.~\cite{Gunn:2024xaq}, we calculate the {\em local} mean free path $\lambda(E,T(r))$ of a photon with energy $E$ using the Breit-Wheeler cross-section calculated over a black body spectrum of temperature $T(r)$ (see App. A for more details). We then estimate the corresponding optical depth of a hotspot of radius $r_{\rm HS}$ for photons of energy $E$ as
\be
{\rm OD}^\gamma(E) \equiv \int_{0}^{r_{\rm HS}}\frac{dr}{\lambda(E,T(r))}\,,
\ee
and the corresponding probability for the photon to escape as $P^\gamma_{\rm esc}(E)\equiv e^{-{\rm OD}^\gamma(E)}$.
We then account for the shielding effect of the hot spot on the photon emission rate by multiplying the flux of secondary photons $dN/(dEdt)|_{\rm sec}$ predicted from PBH evaporation by $P^\gamma_{\rm esc}(E)$.

\vspace{5pt}
\section{Blackbody Radiation\label{sec:BB}}
Throughout the evolution of the hot spot, photons can be emitted locally by the plasma, contributing indirectly to the total flux of photons emitted by the black hole. To account for this effect, we assume the hot spot radiates photons locally like a perfect black body spectrum and estimate the probability of such radiation escaping the hot spot following a similar procedure as above. We define the {\em local} optical depth of the hot spot for photons emitted at radius $r$ as
\begin{equation}
{\rm OD}^\gamma(E,r)\equiv \int_r^{r_{\rm HS}} \frac{d r^{\prime}}{\lambda\left( E, T(r^\prime)\right)}\,,
\label{mfp_av}
\end{equation}
so that the escape probability of such photons from the hot spot is given by 
$P_{\rm esc}^\gamma(E, r)=e^{-{\rm OD}^\gamma(E,r)}$.
Given the emission rate of outgoing photons of energy $E$ emitted by the plasma from a thin shell of radius $r$, $\Phi_{\rm HS}(E,T(r),r)\equiv (E^2/2\pi^2) (e^{E / T(r)}-1)^{-1} 4 \pi r^2$, one can estimate the number of photons escaping the hot spot as 
\begin{equation}
\frac{d\Phi_{\rm HS}}{dr}=\frac{dT}{dr}\cdot\frac{\partial \Phi_{\rm HS}}{\partial T} + \frac{\partial \Phi_{\rm HS}}{\partial r}\,,
\end{equation}
and the emission rate, integrated over the hot spot volume, by
\begin{equation}
\left.\frac{dN}{dtdE}\right|_{\rm HS}=\int_{0}^{r_{\text {HS }}} \left(\frac{d\Phi_{\rm HS}}{dr} \cdot P_{\rm esc}^\gamma(E, r)\right)dr\,.
\end{equation}


\vspace{5pt}
\section{Universal Spectrum and BBN Limits\label{sec:universal}}
After photons escape the hot spot, they undergo electromagnetic cascades by scattering off the Universe's plasma to either pair-produce electrons or scatter through inverse-Compton processes. In a homogeneous Universe, such processes are known to induce a universal spectrum at low energy, function of the energy of the parent particle $E_0$, defined as~\cite{Kawasaki:1994sc,Protheroe:1994dt,Berezinsky:1990qxi}
\be\label{eq:uni}
\left.\frac{dN}{dE}(E_0)\right|_{\rm uni}\!\!\!\!\!\!=K_0\!\left\{\!\!\begin{array}{ll}
     (E/E_X)^{-3/2},&E<E_X  \\
     (E/E_X)^{-2},&E_X\leqslant E\leqslant E_C\\
     0,&E_C< E
\end{array}\right.\!\!\!
\ee
where $K_0\equiv E_0/(E_X^2[2+\ln(E_C/E_X)])$,
 $E_X=m_e^2/(80T)$, and $E_C=m_e^2/(22T)$. In the context of PBH evaporation, each Hawking emitted photon of energy $E_0$ is considered to lead to such a spectrum before the corresponding low-energy photons eventually dissociate light nuclei formed during BBN. Because this spectrum scales linearly with $E_0$, the total spectrum expected from PBH evaporation scales linearly with the total flux of photons emitted by Hawking radiation,
\be
\left.\frac{dN}{dEdt}\right|_{\rm tot} \!\!\!\!\!= \!\int\!\!\! \left.\frac{dN}{dE_0dt}\right|_{\rm sec}\!\!\left.\frac{dN}{dE}(E_0)\right|_{\rm uni}\!\!\!\!\!\!\!dE_0 \propto \!\!\int \!\!\left.\frac{dN}{dE_0dt}\right|_{\rm sec}\!\!\!\! \!\!E_0dE_0\,.
\ee
Assuming the Universe's temperature remains constant throughout the evaporation process, the prefactor in this equation is time-independent. Thus, this same proportionality relation holds after integration over time. The effect of the hot spot on the low-energy photon spectrum thus corresponds to an overall rescaling of the emission flux calculated in a homogeneous universe, up to a transfer function that only depends on the initial mass of the PBH
\be\label{eq:transfer}
\mathcal T(M)\equiv \int \!\!\left.\frac{dN}{dEdt}\right|_{\rm sec}\!\!\!\!\! EP^\gamma_{\rm esc}(E)dE \left[\displaystyle \int \left.\frac{dN}{dEdt}\right|_{\rm sec}\!\!\!\!\!EdE\right]^{-1}\!\!\!\!\!\!\!\!\,.
\ee
In the following, we will use this transfer function as a proxy to estimate the hot spot's impact on all existing constraints from photodissociation.
\vspace{5pt}
\section{Results\label{sec:results}}
\begin{figure}
    \centering
    \includegraphics[width=\linewidth]{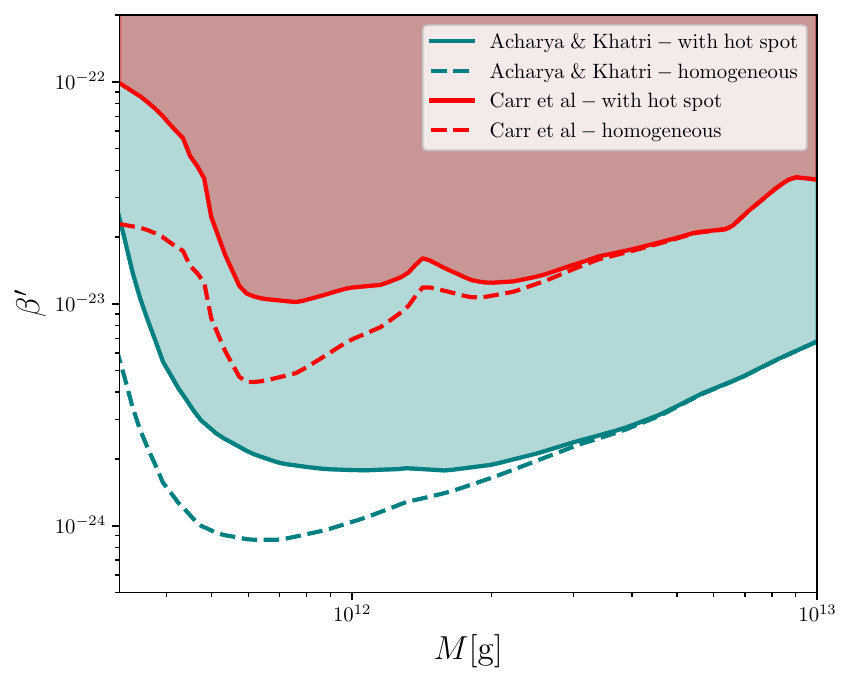}
    \caption{\label{fig:flux}\footnotesize BBN limits on PBHs as derived in~\cite{Acharya_2020} and~\cite{Carr:2020gox} when considering (plain lines) or ignoring (dashed lines) the effect of the hot spot.}
\end{figure}
\begin{figure}
    \centering
    \includegraphics[width=\linewidth]{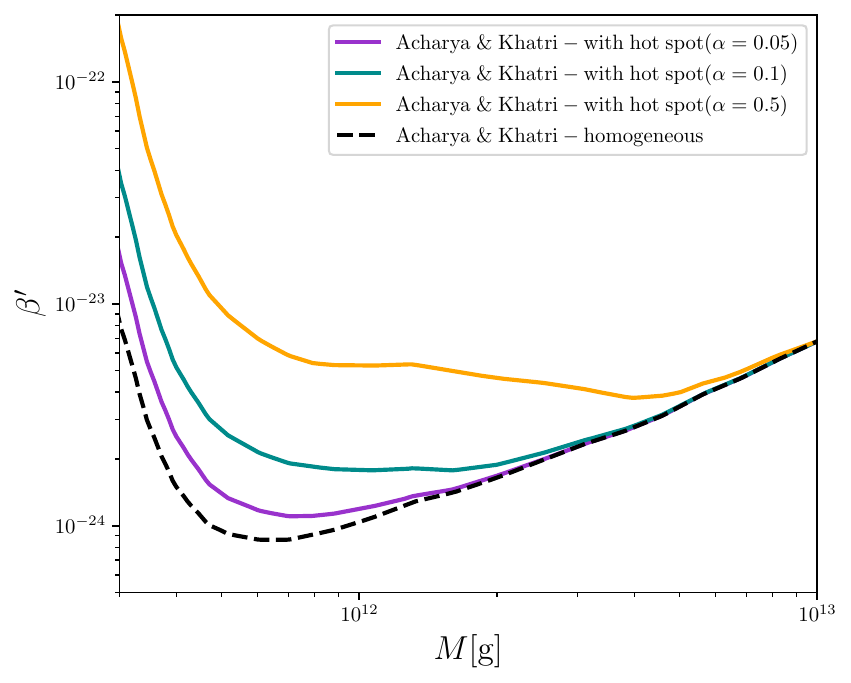}
    \caption{\label{fig:alpha}\footnotesize Variation of the BBN limit on PBHs from Ref.~\cite{Acharya_2020} for varying values of $\alpha$.}
\end{figure}

To track the time evolution of the PBH mass, hot spot profile, and emission rate of secondary photons, we used the software {\tt BlackHawk}~\cite{Arbey:2019mbc, Arbey:2021mbl}. In Fig.~\ref{fig:fluxBlackHawk}, dashed lines represent the fluxes obtained in the absence of hot spots (dashed lines) for photons from Hawking radiation (teal lines) and the corresponding universal spectrum (purple lines). Including the hot spot (plain lines) leads to a sizeable suppression of both fluxes for low PBH masses up to about $M=10^{13}$g. As is visible from the figure, the spectrum of photons radiated directly by the hot spot (orange lines) is always subdominant compared to the other two. The effect of the transfer function $\mathcal  T(M)$ on existing BBN constraints is exhibited in Fig.~\ref{fig:flux}, using the limits obtained in Refs.~\cite{Acharya_2020} and~\cite{Carr:2020gox} and comparing results including and excluding the effect of the hot spot (plain and dashed lines, respectively) in the range of masses where such limits are relevant. One can see from Fig.~\ref{fig:flux} that the effect of the hot spot is sizeable for masses $M\lesssim 4\times 10^{12}\mathrm{g}$, whereas it becomes irrelevant for larger masses. This is due partly to the fact that, for larger PBH masses, the hot spot core temperatures approach the threshold $T_{\rm c}\sim m_e$ where one cannot trust the results exposed in Sec.~\ref{sec:hotspot} anymore, as the thermal bath is mainly made of inert photons. Because we have restricted the effect of the hot spot accordingly, effectively ignoring its existence once its temperature falls below the electron mass, this also means that the escape probability is effectively set to one in that regime. Nonetheless, we stress that the particulars of the hot spot temperature profile strongly rely on the choice of coupling constant $\alpha$ that is assumed to be universal in this paper. In Fig.~\ref{fig:alpha}, we vary this coupling constant, leading to sizeable variations of the corresponding BBN constraints, and of the range of masses over which the effect of the hot spot becomes relevant. In particular,  larger coupling constants (typically at play around the QCD phase transition) lead to a more stringent effect of the hot spot, suggesting that a refined analysis of the exact hot spot temperature profile that forms around PBHs during their evaporation is required to obtain accurate BBN limits in that context.
\vspace{5pt}
\section{Discussion and Conclusion}\label{sec:conclu}
In this paper, we have explored the possibility that the hot spots that surround PBHs when they evaporate affect their capacity to destroy or overproduce light nuclei that formed during BBN. We focused on the case of photodissociation that is particularly relevant in the  mass range $10^{11}\mathrm{g}\lesssim M\lesssim 10^{13}\mathrm{g}$. We investigated two possible ways through which hot spots may have an impact on photodissociation: First, we evaluated the capacity of the hot spot to {\em shield} the emission of photons emitted by PBHs, by calculating the optical depth of the hot spot to the spectrum of photons emitted via Hawking radiation subsequent showering processes. Second, we estimated the fraction of photons that are reprocessed to lower energy via diffusion within the hot spot and emitted occasionally by the hot spot. We finally estimated the universal spectrum of low-energy photons that are reprocessed through scattering off the thermal plasma outside the hot spot, and compared the flux we obtain when ignoring or accounting for the presence of the hot spot. We found that whereas its effect is relatively minor for $M\gtrsim 3\times 10^{12}\mathrm{g}$, the hot spot can drastically affect the flux of low energy photons emitted by the PBH in the MeV-GeV range for PBHs with masses $\mathcal O(10^{11}-10^{12})\mathrm{g}$. Moreover, we observed that the flux of photons emitted directly by the hot spot at low energy remains negligible for all relevant PBH masses.

Before concluding, a few comments regarding the reliability of some of the results presented in this work are in order. First of all, the temperature profile of the hot spot from Ref.~\cite{He:2022wwy} that we used was initially derived assuming that PBHs radiate in a plasma whose temperature is above the QCD phase transition scale. This assumption allowed the authors of Ref.~\cite{He:2022wwy} to consider the case of gluon scattering to obtain ansatzes for the energy deposition and diffusion rates that considerably simplify the discussion. Although a thorough analysis of the hot spot temperature profile would be required at lower temperatures---which we let for future work---we stress that below the QCD scale,  dimensional arguments and a quick comparison of the LPM rates at the typical splitting energy scale~\cite{Mukaida:2022bbo} shows that the aforementioned ansatzes may still hold for electroweak processes, supporting the use high-energy hot spot profile in a purely electroweak plasma. 

To calculate the effect of the hot spot on the photon flux emitted by PBHs, we have assumed that the showering subsequent to primary Hawking emission happened on scales comparable to the hot spot core radius. We emphasize that this radius is obtained by matching the typical first-splitting mean free path of the emitted primary particles with the typical distance over which the thermal bath can thermalise. It is thus a reasonable assumption to consider that splitting, and thus showering, takes place on similar distances. However, in principle, a thorough examination---which is far beyond the scope of this article---of how the temperature gradient affects particle showering and the subsequent spectrum of secondary Hawking emission products expected from PBH evaporation would be required, potentially unveiling surprising results that significantly differ from what the interpolated Pythia simulations used by {\tt BlackHawk} provide.

Finally, we would like to emphasize that the results we have derived, using for simplicity the universal flux of Eq.~\eqref{eq:uni} as the basic ingredient to derive BBN constraints, are conservative. Indeed, this spectrum happens to be underestimated for photons with energies slightly below the pair-creation threshold but above the binding energy of light nuclei, corresponding to the photodissociation threshold. In that energy range (particularly relevant for PBHs with masses $\mathcal O(10^{12}-10^{13})\mathrm{g}$) the actual flux of low energy photons is expected to dominate over the expected universal flux calculated from Eq.~\eqref{eq:uni}~\cite{Poulin_2015,Luo_2021}. Obtaining the exact spectrum accounting for this effect requires solving a Boltzmann equation for the emitted photons~\cite{Luo_2021}, a level of sophistication that we keep for future work, given that the few aforementioned simplifications used in this work may already affect the results to a similar degree.

The evaporation of PBHs and its effects upon cosmology and particle physics have, for a long time, been treated in a homogeneous and isotropic manner. Although recent works suggested otherwise~\cite{Gunn:2024xaq,Hamaide:2023ayu,Das:2021wei,He:2022wwy,He:2024wvt}, the fact that localisation effects may occur despite the large number of evaporating PBHs per Hubble patch remains relatively unexplored in the literature. With this work, we hope to open the path to a new avenue of PBH phenomenology that aims to understand the exact effect of PBH evaporation on the surrounding universe.

\section*{Acknowledgment}
The authors would like to thank J. Gunn, Y.F. Perez-Gonzalez, and J. Turner for inspiring discussions throughout the realisation of this work as well as Kyohei Mukaida for valuable inputs regarding the LPM effect in the electroweak sector. The work of M.~Fairbairn and L.~Heurtier is supported by the U.K.\ Science and Technology Facilities Council (STFC) 
under Grant ST/X000753/1. C.~Altomonte is supported by an NMES faculty grant.

~
\newline~
\vspace{10pt}
\appendix

\section{Hot Spot Temperature Profile}\label{app:photonsMFP}
Within this framework, the authors of Ref.~\cite{He:2022wwy} obtained the following description of the surrounding plasma: For PBHs with masses 
$M \gtrsim M_\star$, where $M_\star \equiv  0.8\mathrm{g}\left(\frac{\alpha}{0.1}\right)^{-\frac{11}{3}}$,
the temperature profile is a piecewise function that features three different regimes: 
\be\label{eq:profile}
T(r) = T_c\times\left\{\begin{matrix}
    1\,, &\quad  r\leqslant r_{\rm c} \\
    \left(\frac{r_{\rm c}}{r}\right)^{3}\,,&\quad r_{\rm c} < r \leqslant r_{\rm diff}\\
    \left(\frac{r_{\rm c}}{r_{\rm diff}}\right)^{3}\left(\frac{r_{\rm diff}}{r}\right)^{7/11}\,,&\quad r_{\rm diff} < r
\end{matrix}\right.
\ee
where we define \vspace{-7pt}
\bea
 \!\!r_{\rm c}&\approx & 8\cdot 10^{8}\ r_H  \left(\frac{\alpha}{0.1}\right)^{-6} \left(\frac{g_{\star}(T_H)}{g_\star(T_{\rm c})}\right)^{-1}\!\!\!\!,\nonumber\\
 \!\!r_{\rm diff}&\approx & 6\cdot  10^{19}\ r_H \nonumber\\
 &\!\times\!&\!\!\left(\frac{\alpha}{0.1}\right)^{-\frac{8}{5}}\!\! \left(\frac{106.75}{g_{\star}(T_H)}\right)^{\frac{4}{5}}\!\!\left(\frac{g_\star(T_{\rm diff})}{106.75}\right)^{\frac{1}{5}}\!\!\left(\frac{M}{10^9\mathrm{g}}\right)^{\frac{6}{5}}\!\!\!\!\,,\nonumber\\
 \eea
 \vspace{-7pt}
 and
 \begin{eqnarray}\label{eq:plateau}
 T_{\rm c} &\approx& 2\times 10^{-4}T_H\left(\frac{\alpha}{0.1}\right)^{\frac{8}{3}} \left(\frac{g_{\star}(T_H)}{g_\star(T_{\rm c})}\right)^{\frac{2}{3}}\,,
 \end{eqnarray}
 where $g_\star(T)$ denotes the number of relativistic degrees of freedom at temperature $T$, $T_H$ denotes the Hawking temperature of the BH, and $T_{\rm diff}\equiv T_{\rm c}(r_{\rm c}/r_{\rm diff})$. 
 In practice, the PBH mass is time-dependent and these expressions for the radii $(r_{\rm c}, r_{\rm diff})$ and the temperature $T_{\rm c}$ need to be evaluated dynamically. Eventually, when the mass reaches the point where $M=M_\star$, diffusion becomes slower than both energy deposition and evaporation, and the energy radiated during the end stage of Hawking evaporation is not redistributed throughout the profile efficiently. Nevertheless, it is argued in Ref.~\cite{He:2022wwy} that this contribution is negligible compared to the energy density accumulated in the profile up to that point.

 At large radii, the profile~\eqref{eq:profile} cannot be trusted indefinitely. For a PBH with initial mass $M_i$, the radius $r_{\rm max}\equiv r_{\rm diff}(M_i)$ sets a maximum distance over which particles cannot diffuse within the lifetime of the BH. When PBHs evaporate, the temperature in the Universe $T_{\rm ev}$ sets another limit to the radius of the profile, as, in practice, one finds that $T_{\rm ev}\gtrsim T(r_{\rm max})$.

\section{Photon Mean Free Path}\label{app:photonsMFP}
Ref. \cite{carmona2022modificationmeanfreepath} considers the mean free path of energetic photons $\gamma$ scattering off background photons $\gamma_b$ due to  the pair production process $\gamma+\gamma_b \rightarrow e^{+}+e^{-}$. 

The threshold energy $\varepsilon^{\text {thr }}$ needed by the background photon $\gamma_b$ as a function of the energy $E$ of the incident photon $\gamma$ and the angle $\theta$ formed by both photons is given by 
\begin{equation}
\varepsilon^{\mathrm{thr}}=\frac{2 m_e^2}{E(1-\cos \theta)}\,.
\end{equation}
The mean-free-path of the incident photon is given by 
\begin{equation}
\frac{1}{\lambda_\gamma(E)}=\frac{1}{8 E^2} \int_{4 m_e^2}^{\infty} d s s \sigma_{\gamma \gamma}(s) \int_{s / 4 E}^{\infty} d \varepsilon \frac{n_\gamma(\varepsilon)}{\varepsilon^2} .
\end{equation}
where $\sigma_{\gamma \gamma}$ is the Breit-Wheeler cross section (where $\alpha$ denotes here the usual electroweak coupling constant, as opposed to the effective coupling constant used in the rest of the paper)
\begin{equation}
\sigma_{\gamma \gamma}(E, \varepsilon, \theta)=\frac{2 \pi \alpha^2}{3 m_e^2} W(\beta)\,,
\end{equation}
with 
\begin{equation}
W(\beta)=\left(1-\beta^2\right)\left[2 \beta\left(\beta^2-2\right)+\left(3-\beta^4\right) \ln \frac{1+\beta}{1-\beta}\right]\,,
\end{equation}
where $\beta$ is the electron and positron velocity in the centre-of-mass frame
\begin{equation}
\beta(\varepsilon, E, \theta)=\sqrt{1-\frac{2 m_e^2}{\varepsilon E(1-\cos \theta)}}=\sqrt{1-\frac{4 m_e^2}{s}}\,.
\end{equation}
and can be rewritten as a function of the relativistic invariant $s$, and $n_\gamma(\varepsilon)$ is the spectral density given by 
\begin{equation}
n_\gamma(\varepsilon)=(\varepsilon / \pi)^2\left(e^{\varepsilon / k T_0}-1\right)^{-1}\,.
\end{equation}
\vfill

\bibliographystyle{apsrev4-1}
\bibliography{main.bib}

\end{document}